\documentclass[pdflatex,sn-basic,Numbered,iicol]{sn-jnl} 

\usepackage[numbers]{natbib}
\usepackage{graphicx}%
\usepackage{multirow}%
\usepackage{amsmath,amssymb,amsfonts}%
\usepackage{amsthm}%
\usepackage{mathrsfs}%
\usepackage[title]{appendix}%
\usepackage{xcolor}%
\usepackage{textcomp}%
\usepackage{manyfoot}%
\usepackage{booktabs}%
\usepackage{algorithm}%
\usepackage{algorithmicx}%
\usepackage{algpseudocode}%
\usepackage{listings}%
\usepackage{glossaries}
\usepackage{subcaption}

\newacronym{itr}{ITR}{Image-Text Retrieval}
\newacronym{cmr}{CMR}{Cross-Modal Retrieval}
\newacronym{fg}{FG}{Fine-Grained}
\newacronym{cg}{CG}{Coarse-Grained}
\newacronym{cmh}{CMH}{Cross-Modal Hashing}
\newacronym{vlp}{VLP}{Vision-Language Pretraining}

\begin{document}

\title[Article Title]{FiCo-ITR: bridging fine-grained and coarse-grained image-text retrieval for comparative performance analysis}

\author*[]{\fnm{Mikel} \sur{Williams-Lekuona}}\email{m.williams@lboro.ac.uk}

\author[]{\fnm{Georgina} \sur{Cosma}}\email{g.cosma@lboro.ac.uk}

\affil[]{\orgdiv{Computer Science Department}, \orgname{Loughborough University}, \orgaddress{\street{Epinal Way}, \city{Loughborough}, \postcode{LE11 3TU}, \state{Leicestershire}, \country{United Kingdom}}}

\abstract{In the field of Image-Text Retrieval (ITR), recent advancements have leveraged large-scale Vision-Language Pretraining (VLP) for Fine-Grained (FG) instance-level retrieval, achieving high accuracy at the cost of increased computational complexity. For Coarse-Grained (CG) category-level retrieval, prominent approaches employ Cross-Modal Hashing (CMH) to prioritise efficiency, albeit at the cost of retrieval performance. Due to differences in methodologies, FG and CG models are rarely compared directly within evaluations in the literature, resulting in a lack of empirical data quantifying the retrieval performance-efficiency tradeoffs between the two. This paper addresses this gap by introducing the \texttt{FiCo-ITR} library, which standardises evaluation methodologies for both FG and CG models, facilitating direct comparisons. We conduct empirical evaluations of representative models from both subfields, analysing precision, recall, and computational complexity across varying data scales. Our findings offer new insights into the performance-efficiency trade-offs between recent representative FG and CG models, highlighting their respective strengths and limitations. These findings provide the foundation necessary to make more informed decisions regarding model selection for specific retrieval tasks and highlight avenues for future research into hybrid systems that leverage the strengths of both FG and CG approaches.}


\maketitle

\section{Introduction}\label{}

\gls*{cmr} involves using one type of data, such as text, to search for another type of data, such as images. Unlike general multi-modal tasks, \gls*{cmr} specifically focuses on bridging the gap between different modalities to enable retrievals across modalities. \gls*{cmr} has gained prominence over the past decade due to its success in various applications, including e-commerce~\citep{Ma2022EICLIPEI, Zhang2021FashionFM}, content-based retrieval~\citep{nakatsuka2023content, gong2023neural}, video surveillance~\citep{yang2024detection}, and recommendation systems~\citep{Truong2021ExploringCU}. When the retrieval task specifically involves images and text, it is referred to as \gls*{itr}. There are two distinct subfields within \gls*{itr}: \gls*{fg} and \gls*{cg} \gls*{itr}.

\gls*{fg} \gls*{itr} aims to find instance-level matches, retrieving the image that directly corresponds to a detailed text query, and vice versa. State-of-the-art \gls*{fg} methods employ large-scale \gls*{vlp} followed by retrieval fine-tuning, often through the use of contrastive learning techniques~\citep{gan2022vision}.

\begin{figure*}[!htpb]
\centering
\includegraphics[width=0.98\linewidth]{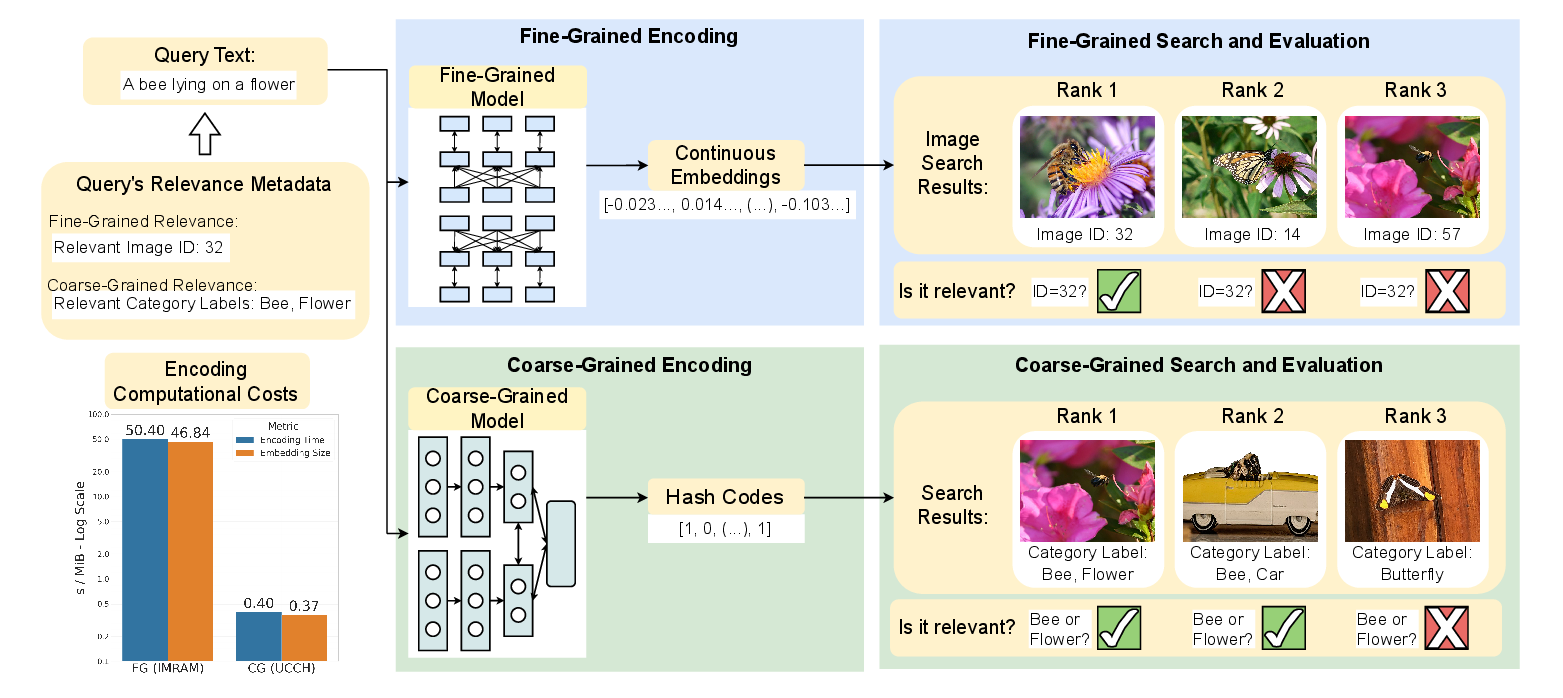}
\caption{Comparison of Fine-Grained (FG) and Coarse-Grained (CG) Image-Text Retrieval approaches. FG search uses continuous embeddings, aiming to find the retrieval sample that specifically corresponds to the query sample. Under evaluation conditions, this involves finding the retrieval sample with the same ID as the query sample. CG search employs bitwise hash codes to find retrieval samples that are more broadly relevant to the query instead of exact matches. During evaluation, this involves finding any retrieval sample with at least one matching category label relative to the query's relevant category labels. The broader search criteria of CG search allows for more efficient computational costs, as seen in the comparison of encoding time and embedding storage costs of two representative FG and CG models (IMRAM~\citep{chen2020imram} and UCCH~\citep{hu2022unsupervised})}
\label{fig:fgvscg}
\end{figure*}

\gls*{cg} \gls*{itr} focuses on category-level retrieval, where the retrieved samples should broadly belong to the semantic category which the query is searching for, rather than aiming for specific exact matches. State-of-the-art \gls*{cg} methods implement Cross-Modal Hashing~\citep{luo2023survey}, which train hash functions to map image and text samples onto a common Hamming subspace for efficient bitwise similarity comparisons. Figure \ref{fig:fgvscg} illustrates the different search criteria between \gls*{fg} and \gls*{cg} search approaches.

Despite sharing the same overarching retrieval task, the subfields of \gls*{fg} and \gls*{cg} \gls*{itr} have evolved independently with limited integration, leading to the following three major challenges: 1) Because of methodological differences between the two subfields, obtaining informative quantitative results which are directly comparable is a non-trivial task. 2) Researchers in one subfield rarely benchmark against works from the other in a comprehensive manner. Although \gls*{fg} and \gls*{cg} methods have been comparatively surveyed within the literature~\citep{zhu2023cross}, direct empirical comparative evaluations of recent representative models are lacking. 3) Traditional \gls*{itr} benchmark datasets such as Flickr30K~\citep{plummer2015flickr30k} (1K sample test set) and MS-COCO~\citep{lin2014microsoft} (5K sample test set) are small compared to real-world applications. Therefore, these benchmark datasets do not offer a comprehensive understanding of model performance on large-scale data, potentially skewing perceived retrieval performance-efficiency trade-offs. 

The lack of standardised evaluation methodologies and empirical comparisons between \gls*{fg} and \gls*{cg} methods has hindered the field's understanding of their relative strengths and limitations in real-world scenarios. This understanding is necessary for informing model selection decisions and identifying opportunities for hybrid approaches that could leverage the advantages of both methodologies. To address these critical gaps, this paper makes the following contributions:

\begin{enumerate}
    \item The implementation and distribution of the \texttt{FiCo-ITR} library and unified toolkit that bridges the methodological differences between the \gls*{cg} and \gls*{fg} \gls*{itr} subfields. This contribution addresses the challenge of establishing fair comparisons by implementing carefully selected tasks, datasets, and metrics that accommodate the distinct characteristics of each subfield while maintaining evaluation consistency. The library is made available in the Python Package Index (PyPI) and GitHub, enabling researchers to conduct standardised evaluations across both subfields, facilitating more informed comparisons and potential cross-pollination of ideas between the two approaches.
    
    \item A systematic empirical evaluation that quantifies the trade-offs between \gls*{cg} and \gls*{fg} approaches across multiple aspects of performance: Recall, precision, encoding time, storage costs, query-time attention costs and similarity search time. This evaluation provides the first comprehensive empirical basis for understanding the relative strengths and limitations of representative approaches from both subfields, addressing the current lack of direct comparative analysis in the literature.
    
    \item Scalability experiments that evaluate model performance using incrementally larger retrieval sets, quantifying the computational costs of large-scale \gls*{itr}. These experiments reveal critical trade-offs between retrieval performance and computational efficiency that only become apparent at scale, providing practical guidance for real-world deployment decisions.
\end{enumerate}

Section~\ref{sec:methods} provides an overview of \gls*{fg} and \gls*{cg} \gls*{itr}, highlighting key trends and major works. Section~\ref{sec:prop} outlines the implementation of the proposed \texttt{FiCo-ITR} library. Section~\ref{sec:methodology} details the experiment methodology for the conducted evaluations. Section~\ref{sec:comp} employs the library to jointly evaluate representative \gls*{fg} and \gls*{cg} models. Section~\ref{sec:discussion} discusses the key findings, providing recommendations for the use-cases of \gls*{fg} and \gls*{cg} \gls*{itr}. Section~\ref{sec:conclusion} concludes the paper.

\section{Related works}\label{sec:methods}

\subsection{Fine-grained image-text retrieval}\label{sec:cmm}

\gls*{fg} methods aim to map visual and textual information to a joint space where relevant samples are aligned at the instance level. The alignment of samples is achieved through learning shared feature representations from image-text pairs during training. Establishing the joint space enables retrieval across modalities, where the relative distance of items in the shared space determines their relevance. While \gls*{fg} methods share the same fundamental principles as \gls*{cg} ones, they differ in the level of scrutiny with which the processed samples are analysed. Specifically, \gls*{fg} methods focus on low-level features and object-level relationships to achieve an in-depth understanding of the scene within a given sample. The format of the encoded samples is typically real-valued continuous embeddings. These embeddings enable calculating the similarity between samples using vector distance measurements such as Cosine distance~\citep{faghri2017vse++}.

\textbf{Traditional fine-grained methods.} Early deep learning-based \gls*{fg} methods typically employed Long Short-Term Memory (LSTM) networks for text encoding and convolutional neural networks (CNNs) such as AlexNet and VGGNet for image feature extraction~\citep{kiros2014unifying}. VSE++~\citep{faghri2017vse++} built on this approach by experimenting with VGG19~\citep{simonyan2014very} and ResNet152~\citep{he2016deep} for image encoding along with a GRU-based text encoder and implementing hard negative mining and reranking loss functions. SCAN~\citep{lee2018stacked} achieved a breakthrough in performance through its proposed bottom-up attention module aligning image regions and words based on Faster R-CNN object detection~\citep{ren2015faster}. CAMP~\citep{wang2019camp} aggregates salient messages between image regions and words via attention to handle negative pairs before directly predicting matching scores. VSRN~\citep{li2019visual} introduced hierarchical reasoning to capture region-concept interactions and was later upgraded to VSRN++~\citep{li2022image} using BERT~\citep{devlin2019bert} text features. KASCE~\citep{shi2019knowledge} expands image concepts using ``common-sense'' relationships from scene graphs and selects the most relevant expansions. CAAN~\citep{zhang2020context} employs context-aware attention to selectively attend to fragments based on inter- and intra-modal semantics. IMRAM~\citep{chen2020imram} progressively aligns fragments through a recurrent attention unit to capture different semantics at each step, alongside a memory unit to accumulate cues. 

\textbf{Fine-grained vision-language pretraining methods.} Following the success of the attention mechanism in BERT~\citep{devlin2019bert}, the transformer architecture has been extensively adapted for vision-language tasks. Moreover, first pioneered by the model ViLT~\citep{kim2021vilt}, state-of-the-art performance is achieved by using transformer encoder stacks end-to-end, without the need for additional pre-processing steps such as object detection feature extraction predominant in traditional \gls*{fg} models. This shift most notably involves leveraging pretraining on large-scale data to obtain task-agnostic features for subsequent fine-tuning onto downstream tasks such as \gls*{itr}. Initial \gls*{vlp} methods are categorised into two main encoder architectures: Dual-encoder and fusion-encoder architectures. Dual-encoders, such as ALIGN~\citep{jia2021scaling} and BEIT-3~\citep{wang2023image}, employ separate image and text encoders, allowing the independent computation and offline storage of embeddings for each modality. This approach eliminates the need to compute embeddings at query time. Fusion-encoder models, such as UNITER~\citep{chen2020uniter} and METER~\citep{dou2022empirical}, enhance interactivity between modalities by employing a unified encoder that jointly processes image and text inputs. Although the fusion encoder approach may potentially lead to higher-quality embeddings, the embeddings must be computed at query time due to the requirement to process all possible query/retrieval sample combinations jointly.  Building on these two approaches, models such as X2-VLM~\citep{zeng2023x} and BLIP-2~\citep{li2023blip} adopt a hybrid methodology. These hybrid methods first leverage a dual-encoder step to independently compute embeddings to boost efficiency. Then, to further improve the embedding quality, they employ a fusion-encoder reranking step that allows for modality interaction. Specifically, the top candidates retrieved from the initial dual-encoder step are reranked using the fusion-encoder during query time.

\textbf{Efficiency in \gls*{fg} \gls*{itr}.} To address efficiency challenges inherent to \gls*{fg} \gls*{itr}, recent approaches have aimed to optimise inference speeds by proposing lightweight implementations of the dual-encoder architecture. LightningDOT~\citep{sun2021lightningdot} tackles this by simplifying pre-training tasks, maximising the amount of computations that can be done offline, and promoting the use of lightweight encoders. VLDeformer~\citep{zhang2022vldeformer} proposes a two-stage retrieval process, first using a transformer learning stage followed by more efficient indexing-based retrieval in the second stage. HiVLP~\citep{chen2022hivlp} uses coarse-grained screening as a first step, followed by a fine-grained rerank in the second step. However, rather than a hash-based approach, HiVLP generates features for the coarse step using early layers of the transformer stack, whereas fully inferred features from later layers are used for the fine-grained rerank.

\subsection{Coarse-grained image-text retrieval}\label{sec:cmh}

\gls*{cg} methods, similarly to \gls*{fg} methods, also aim to learn joint visual and textual representations. However, in contrast to the instance-level alignment of \gls*{fg} methods, \gls*{cg} methods aim to align relevant samples at a broader semantic category level. By using a broader criterion, \gls*{cg} methods can place more emphasis on computational efficiency. The most prominent approach within \gls*{cg} \gls*{itr} is \gls*{cmh}~\citep{luo2023survey}, which is primarily characterised by the use of bit hash codes to represent their encoded data. The use of hash codes aims for lower storage costs and faster retrieval speeds, due to the bit hash format being inherently lightweight and the associated bitwise operations being less computationally complex than continuous embedding operations.  Recent \gls*{cmh} methods use deep neural networks to learn hash functions that map image and text samples to binary hash codes. During retrieval, the hash codes are compared using Hamming distance, an efficient similarity measure for bit strings which counts the number of differing bits between two equal-length bit hash codes.

\textbf{Supervised cross-modal hashing.} Supervised \gls*{cmh} methods leverage multi-category labelling to train the hash function for each modality. DCMH~\citep{jiang2017deep} pioneered this approach by proposing an end-to-end deep learning \gls*{cmh} framework. Leveraging Generative Adversarial Networks (GAN), SSAH~\citep{li2018self} implements label information as network input to strengthen category alignment in the hash space. AGAH~\citep{gu2019adversary} uses label information directly in its loss function, implementing a multi-labeling map. DADH~\citep{bai2020deep} adopts a weighted cosine triplet-margin constraint for ranking-based similarity preservation. DCHUC~\citep{tu2020deep} introduces a four-step iterative optimisation process that allows simultaneous learning of unified hash codes for database samples and modality-specific hashing functions for unseen queries. DCHMT~\citep{tu2022differentiable} employs a dual transformer tower network and a differentiable hashing module, enabling location-aware image encoding and continuous, gradient descent-optimised modality representation. LDSAH~\citep{li2023label} integrates label-wise semantic alignment with a dissimilarity-penalising strategy using a combination of Jensen–Shannon divergence loss and attention-driven sample re-weighting.

\textbf{Unsupervised cross-modal hashing. } Unsupervised \gls*{cmh} methods use image-text pair coupling information to learn the modality hash functions, avoiding the reliance on category labelling information. This makes unsupervised \gls*{cmh} methods more analogous to typical \gls*{fg} methods, which also do not rely on labelling. Unsupervised UGACH~\citep{zhang2018unsupervised} uses GANs to exploit the underlying manifold structure of cross-modal data with a max-margin ranking loss. UDCMH~\citep{wu2018unsupervised} constructs self-taught deep hash functions by minimising quantisation errors while preserving nearest and farthest neighbourhood relationships. DJSRH~\citep{su2019deep} uses a joint semantics affinity matrix to combine neighbourhood relations from different modalities for improved quantisation and batch-wise training efficiency. DSAH~\citep{yang2020deep} employs a semantic-alignment loss function with auto-encoder-based hash code generation. JDSH~\citep{liu2020joint} proposes a similarity decision and weighting approach which uses a threshold-based weighting scheme to increase the discrimination of hash codes. DGCPN~\citep{tu2020deep} uses a graph-based framework with nodes representing related data and employs a conditional probability model to evaluate the coherence between neighbouring nodes. ADV~\citep{li2021efficient} adopts fine-grained learning objectives for a two-step instance-level retrieval task where shorter hash code embeddings perform initial screening for subsequent reranking using longer hash codes. UCCH~\citep{hu2022unsupervised} uses contrastive learning with a momentum optimiser and introduces a cross-modal ranking learning loss to address binary-continuous relaxation challenges and mitigate the impact of false-negative pairs.

\section{Proposed library and toolkit}\label{sec:prop}

\begin{figure*}[ht]
\centering
\includegraphics[width=0.98\linewidth]{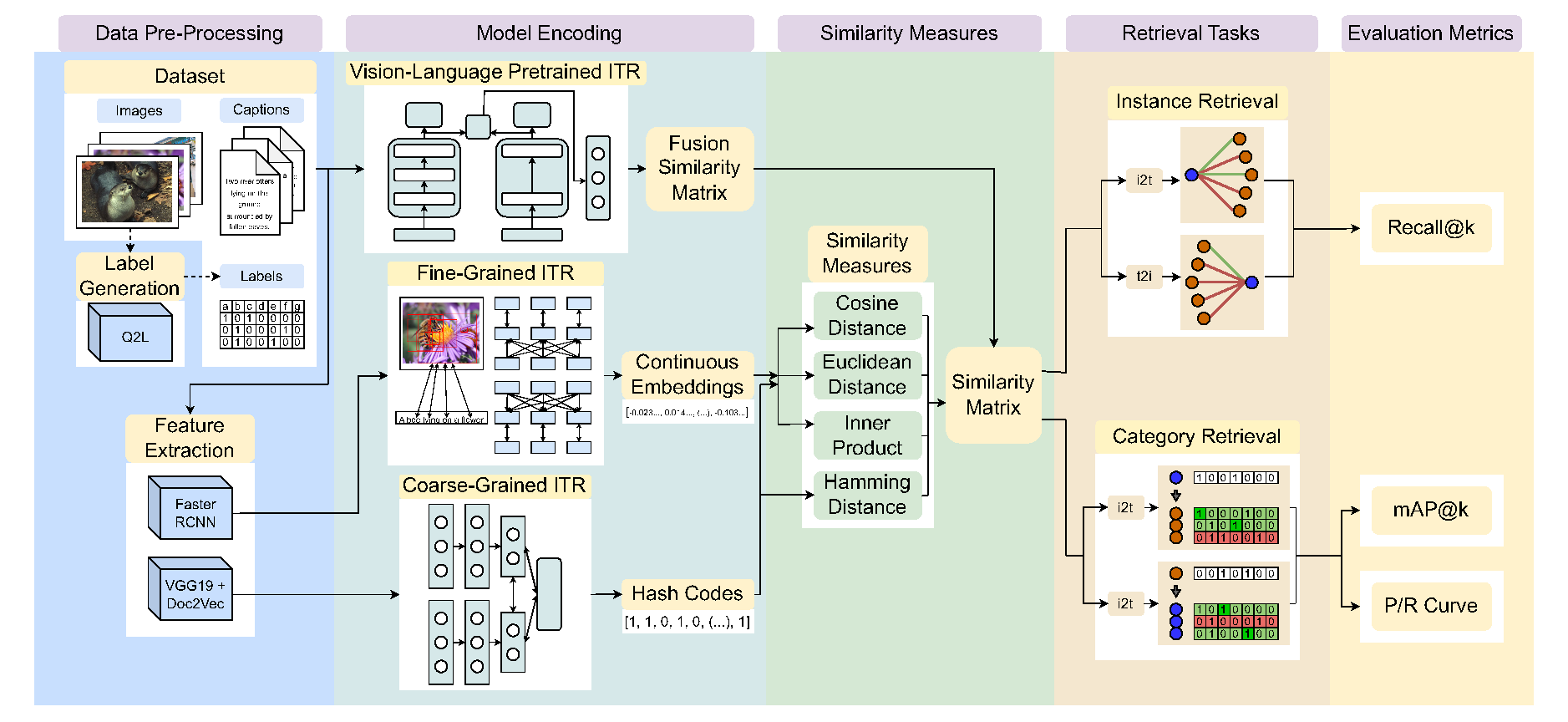}
\caption{An extendable framework of the FiCo-ITR library and toolkit. The pipeline consists of five main components: 1) Data Pre-Processing, which standardises dataset handling and offers optional label generation via Query2Label (Q2L)~\citep{liu2021query2label} for unlabeled datasets; 2) Model Encoding, supporting embeddings in the form of binary hash codes and continuous embeddings, as well as direct similarity matrices; 3) Similarity Measures, implementing four distance measures for uniform similarity calculation; 4) Retrieval Tasks, implementing instance- and category-level retrieval for both (i $\rightarrow$ t) and (t $\rightarrow$ i); and 5) Evaluation Metrics, reporting Recall@K for instance-level retrieval and mAP@K with P/R curves for category-level retrieval}
\label{fig:i2tretrieval}
\end{figure*}

We propose \texttt{FiCo-ITR}, a comprehensive library and toolkit that unifies the evaluation of the \gls*{fg} and \gls*{cg} \gls*{itr} subfields, facilitating direct comparisons for both image-to-text (i $\rightarrow$ t) and text-to-image (t $\rightarrow$ i) retrieval tasks. As illustrated in Figure~\ref{fig:i2tretrieval}, the library's framework comprises five key components: 1) Data Pre-Processing, 2) Model Encoding, 3) Similarity Metrics, 4) Retrieval Tasks, and 5) Evaluation Metrics, forming a complete pipeline for image-text retrieval evaluation.

\textbf{1) Data pre-processing.} To enable both instance-level and category-level retrieval evaluations, the selected benchmark datasets must have full-sentence captions as well as semantic category labelling. For datasets which do not have semantic category labelling, we employ the Query2Label (Q2L)~\citep{liu2021query2label} classifier to label them with synthetic semantic category labels. We adopt pre-processing steps which most closely adhere to each subfield's standard approaches~\citep{hu2022unsupervised, lee2018stacked}: For \gls*{cg} models, \texttt{FiCo-ITR} provides the toolkit to extract 4096D VGG-19 features for images and 300D Doc2Vec features for text. Images are preprocessed by resizing to VGG-19's expected 224×224 pixels and applying standard normalisation (mean=[0.485, 0.456, 0.406], std=[0.229, 0.224, 0.225]). The VGG-19 network, pretrained on ImageNet, has its final three layers removed to obtain the feature vectors. Text features are generated using Doc2Vec trained on the combined training and test captions, with minimum word count thresholding to handle vocabulary sparsity. For traditional \gls*{fg} models, we refer to the Bottom-Up Attention~\citep{Anderson2017up-down} repository for Faster R-CNN to extract 36 region proposals per image with 1024-dimensional features, whereas the raw text is given as input to each model's specific encoder. VLP models process raw data using their respective transformer encoder stacks.

\textbf{2) Model encoding interface.} The framework provides standardised interfaces for handling model outputs while maintaining evaluation consistency across different encoding approaches. While the actual encoding process is model-specific and outside the scope of our framework, \texttt{FiCo-ITR} implements pre-encoding dataset management and post-encoding output handling. For dataset management, the toolkit provides functionality to split samples into pre-defined splits along their IDs, ensuring that when different models encode the same test set, they process exactly the same samples in the same order. For output handling, the framework supports three types of encoded outputs: Binary hash codes from \gls*{cg} models (typically 64-bit to 2048-bit lengths), continuous embeddings from \gls*{fg} models (typically 256 to 2048 dimensions), and directly computed similarity matrices from models employing query-time attention. This flexible interface design enables fair comparisons while respecting each model's native encoding process.

\textbf{3) Similarity measures.} Given a set of evaluation embeddings, we implement four similarity measures optimised for different embedding types:

\begin{itemize}
    \item Cosine similarity for continuous embeddings, computed as the normalised dot product between vectors $\mathbf{x}$ and $\mathbf{y}$: $\frac{\mathbf{x} \cdot \mathbf{y}}{\|\mathbf{x}\| \|\mathbf{y}\|}$, with values ranging from [-1, 1]
    \item Euclidean similarity, transformed to a (0, 1] range through $1/(1 + d + \epsilon)$ where $d$ is the Euclidean distance $\sqrt{\sum(\mathbf{x} - \mathbf{y})^2}$ and $\epsilon=10^{-8}$ for numerical stability
    \item Hamming similarity for binary hash codes, computed as the negative normalised count of differing bits: $-\sum(\mathbf{x} \neq \mathbf{y})/n$ where $n$ is the bit length
    \item Inner product similarity $\mathbf{x} \cdot \mathbf{y}$, particularly suitable for embeddings trained with dot-product-based loss functions
\end{itemize}

For traditional binary hash codes (0s and 1s), we implement Hamming distance which calculates the number of differing bits between two hash codes efficiently on CPUs. The library also supports hash codes represented as -1s and 1s, enabling the use of inner product, cosine similarity, and Euclidean distance. This design choice allows users to leverage GPU acceleration for hash code comparisons, potentially boosting performance in GPU-centric environments. For continuous embeddings, all measures except Hamming distance are applicable.

\textbf{4) Retrieval tasks.} Given a similarity matrix, \texttt{FiCo-ITR} implements two retrieval task evaluations: Instance-level retrieval and category-level retrieval. For the instance-level retrieval task, the objective is to search for the retrieval sample which directly corresponds to the query sample. Given an image with multiple captions, for the (i $\rightarrow$ t) task, retrieving any one of the multiple captions of the query image is considered a 100\% recall. For the (t $\rightarrow$ i) task, the specific image corresponding to the query caption must be retrieved for the query to be correct. For the category-level retrieval task, the objective is to retrieve samples that share at least one of multiple semantic category labels with the query sample. This approach results in a broader definition of relevance, where a large amount of samples in the retrieval set may be considered relevant to a given query based on shared category labels.

\textbf{5) Evaluation metrics.} With the retrieval results produced by the retrieval tasks, for instance-level retrieval, we use recall at $k$ ($R@k$) as the primary retrieval performance metric, defined as follows:

\begin{equation}
R@k = \frac{\text{number of relevant items in top } k \text{ results}}{\text{total relevant items}}
\end{equation}

Recall is particularly suitable for instance-level retrieval as it measures whether specific, individual items have been found. For category-level retrieval, we use mean average precision at $k$ (mAP$@k$) as the standard evaluation metric. Precision at $k$ ($P@k$) is defined as follows:

\begin{equation}
P@k = \frac{\text{number of relevant items in top } k \text{ results}}{\text{number of retrieved items}}
\end{equation}

For a single query, AP$@k$ averages the precision values at all available ranks within the top $k$. For a query set, mAP$@k$ computes the mean of the AP$@k$ scores of all the queries within the query set. mAP$@k$ is well-suited for evaluating category-level retrieval where queries typically have a large number of relevant items, as it measures a model's ability to include many relevant results towards the top of a large retrieved set. Additionally, we implement the 11-point precision-recall curve for the category-level retrieval task, which plots interpolated precision at recall levels $R = \{0.0, 0.1, \ldots, 1.0\}$.

\section{Experiment methodology} \label{sec:methodology}

The following is the experiment methodology for the comparative experiments conducted in Section \ref{sec:comp}.

\textbf{Datasets.} We use MS-COCO and Flickr30K as primary benchmark datasets. MS-COCO contains 123\,287 images, each with five human-annotated captions and native multi-label annotations across 80 semantic categories, making it particularly suitable for evaluating both instance-level and category-level retrieval approaches. MS-COCO's adoption as a standard benchmark across both \gls*{fg} and \gls*{cg} communities \citep{gan2022vision, luo2023survey} further enables fair cross-methodology comparisons. Flickr30K comprises 31K images, also with five human-annotated captions per image, focusing on human-object interactions. To enhance the comparability of the results, the toolkit provided through our \texttt{FiCo-ITR} library implements the Karpathy~\citep{karpathy2017deep} split uniformly across all evaluated models for consistency. This split allocates 5K images for testing and 5K for validation in MS-COCO, with the remainder used for training. For Flickr30K, 1K images are designated for testing and validation each, with the remaining 29K samples used for training. Due to Flickr30K not containing semantic category labelling, we employ the label generation module of \texttt{FiCo-ITR} enabled by Q2L~\citep{liu2021query2label} to generate semantic category labelling for it. These generated labels are available for use within the provided \texttt{FiCo-ITR} repository.

\textbf{Model selection.} The model selection process considered both academic impact and architectural diversity while ensuring practical reproducibility. From recent surveys and benchmarks \citep{luo2023survey, gan2022vision}, we identified representative models across the key architectural approaches in \gls*{itr} (as detailed in Table \ref{tab:models}). Our model selection thus aims to span the architectural spectrum: For VLP models, we include the fusion-encoder ViLT~\citep{kim2021vilt}, dual-encoder BEiT-3~\citep{wang2023image}, and hybrid dual+fusion rerank models BLIP-2~\citep{li2023blip} and X-VLM~\citep{zeng2022multi}. Traditional \gls*{fg} models are represented through both local+global attention (IMRAM~\citep{chen2020imram}, SCAN~\citep{lee2018stacked}) and global-only attention (VSRN~\citep{li2019visual}) approaches, while \gls*{cg} models cover the main paradigms: supervised (DADH~\citep{bai2020deep}), unsupervised (UCCH~\citep{hu2022unsupervised}), and quantisation-based (ADV~\citep{li2021efficient}) approaches. Some efficiency-focused VLP models of interest (LightningDot~\citep{sun2021lightningdot}, VLDeformer~\citep{zhang2022vldeformer}, and HiVLP~\citep{chen2022hivlp}) could not be included due to practical constraints such as inaccessible data dependencies or lack of public implementations. To partially address this gap, we implement dual-encoder-only variants of BLIP-2 and X-VLM to promote efficiency ($\text{BLIP-2}_{\text{NF}}$ and $\text{X-VLM}_{\text{NF}}$). Additionally, we evaluate ADV in two configurations: as a \gls*{cg} model with 64-bit hash codes and as a \gls*{fg} model with 2048-bit codes. Table \ref{tab:models} provides a detailed overview of the architectural characteristics and computational requirements across our model selection.

\textbf{Evaluation metrics.} We use the following configurations of the metrics implemented in the \texttt{FiCo-ITR} library for our experiments: Instance-level retrieval performance is assessed using $R@$1, $R@$5, and $R@$10. Additionally, we explore $R@$50, $R@$100, and $R@$200 for the \gls*{cg} models to examine the viability of their use case as first-screening steps where their top-k results are passed to a \gls*{fg} model for further reranking. For category-level retrieval, we employ mAP@10, mAP@100, and mAP@$N$ (where $N$ is the total number of retrieval candidates), offering a comprehensive view of performance across increasing retrieval depths while having several relevant samples available. To further illustrate the trade-off between precision and recall in category-level retrieval, we include 11-point interpolated precision-recall curves.

\begin{table*}[h]
\centering
\begin{tabular}{lllllll} \hline
Model & Approach                   & Q-Time Attn & Dim   & Img Feat    & Txt Feat   & DType \\ \hline
\multicolumn{7}{c}{Fine-Grained Vision-Language Pretrained} \\
BLIP-2  & Dual+Fusion                & Fusion Rerank   & 256D   & 1408x677D    & 768×40D    & f32   \\
BEIT-3  & Dual                       & None          & 768D   & -           & -          & f32   \\
X-VLM   & Dual+Fusion                & Fusion Rerank   & 256D   & 1024×145D    & 768×40D    & f32   \\
ViLT & Fusion & Full Fusion & n/a & 768x217D & 768×40D & f32 \\ \hline
\multicolumn{7}{c}{Fine-Grained} \\
IMRAM   & Local+Global               & Cross-Attn       & 1024D  & 1024×36D     & 1024×70D    & f64   \\
VSRN    & Global                     & None          & 2048D  & -           & -          & f64   \\
SCAN    & Local+Global               & Cross-Attn       & 1024D  & 1024×36D     & 1024×70D    & f64   \\ \hline
\multicolumn{7}{c}{Coarse-Grained} \\
ADV     & Dual                       & None          & 64D/2048D & -         & -          & i8    \\
UCCH    & Dual Hash                  & None          & 64D    & -           & -          & i8    \\
DADH    & Dual Hash                  & None          & 64D    & -           & -          & i8    \\ \hline
\end{tabular}
\caption{Architectural details and computational characteristics of the evaluated image-text retrieval models. For models using query time attention, the image and text features used are shown as these are necessary during query time computation for their respective attention steps}
\label{tab:models}
\end{table*}

\section{Comparative experiments}\label{sec:comp}
\subsection{Instance-level retrieval results}
This experiment aims to empirically assess the comparative performance of \gls*{cg} and \gls*{fg} models in instance-level retrieval tasks. Through the use of the proposed \texttt{FiCo-ITR} library, the instance-level Recall@k evaluation results for the selected models on the Flickr30K and MS-COCO datasets are reported in Table \ref{tab:recall}, with additional results at higher top-k levels for the \gls*{cg} models being reported in Table \ref{tab:recall2}. From these results, the following observations can be made:

\begin{table*}[h]
\centering
\begin{tabular}{lllllllllllll} \hline
\multicolumn{13}{c}{Instance-Level Retrieval Task}\\ \hline
\multicolumn{1}{c}{\multirow{2}{*}{\begin{tabular}[c]{@{}c@{}}\\  \end{tabular}}} & \multicolumn{6}{c}{MS-COCO} & \multicolumn{6}{c}{Flickr30K} \\
\multicolumn{1}{c}{} & \multicolumn{3}{c}{Image to Text} & \multicolumn{3}{c}{Text to Image} & \multicolumn{3}{c}{Image to Text} & \multicolumn{3}{c}{Text to Image} \\
\multicolumn{1}{l}{Recall@k} & \multicolumn{1}{c}{@1} & \multicolumn{1}{c}{@5} & \multicolumn{1}{c}{@10} & \multicolumn{1}{c}{@1} & \multicolumn{1}{c}{@5} & \multicolumn{1}{c}{@10} & \multicolumn{1}{c}{@1} & \multicolumn{1}{c}{@5} & \multicolumn{1}{c}{@10} & \multicolumn{1}{c}{@1} & \multicolumn{1}{c}{@5} & \multicolumn{1}{c}{@10} \\ \hline
& \multicolumn{12}{c}{Fine-Grained Vision-Language Pretrained} \\
BLIP-2 & \textbf{85.4} & \textbf{97.0} & \textbf{98.4} & \textbf{68.2} & \textbf{87.2} & \textbf{92.6} & \multicolumn{1}{r}{\textbf{97.6}} & \multicolumn{1}{r}{\textbf{100.0}} & \multicolumn{1}{r}{\textbf{100.0}} & \multicolumn{1}{r}{\textbf{89.7}} & \multicolumn{1}{r}{\textbf{98.1}} & \multicolumn{1}{r}{\textbf{98.9}} \\
$\text{BLIP-2}_{\text{NF}}$ & 74.3 & 94.2 & 97.4 & 63.5 & 86.1 & 91.8 & 90.8 & 99.6 & 99.9 & 85.7 & 97.6 & 99.1 \\
BEIT-3 & 79.0 & 94.3 & 97.2 & 61.3 & 84.6 & 90.7 & 96.3 & 99.7 & \textbf{100.0} & 86.1 & 97.6 & 98.8 \\
X-VLM & 81.0 & 95.1 & 98.0 & 63.0 & 85.7 & 91.5 & 96.8 & 99.8 & \textbf{100.0} & 86.1 & 97.4 & 98.7 \\
$\text{X-VLM}_{\text{NF}}$ & 71.4 & 91.9 & 96.3 & 54.4 & 81.5 & 88.8 & 90.2 & 99.0 & 99.7 & 78.4 & 95.2 & 97.8 \\
ViLT & 61.6 &	86.3 &	92.7 &	43.0 &	72.7 &	83.1 &	83.8 &	96.8 &	98.6 &	65.2 &	88.7 &	93.6 \\ \hline
& \multicolumn{12}{c}{Fine-Grained} \\
IMRAM & 54.5 & 83.1 & 90.3 & 39.5 & 68.8 & 79.8 & 71.0 & 93.0 & 96.5 & 52.1 & 78.7 & 86.0 \\
VSRN & 50.3 & 79.5 & 87.9 & 37.9 & 58.6 & 79.4 & 70.4 & 89.2 & 93.7 & 53.0 & 77.9 & 85.7 \\
SCAN & 44.9 & 76.7 & 86.7 & 33.7 & 62.8 & 75.2 & 66.7 & 89.3 & 94.0 & 43.1 & 73.4 & 82.2 \\
$\text{ADV}_{\text{2048bit}}$ & \multicolumn{1}{c}{-} & \multicolumn{1}{c}{-} & \multicolumn{1}{c}{-} & \multicolumn{1}{c}{-} & \multicolumn{1}{c}{-} & \multicolumn{1}{c}{-} & 63.7 & 87.1 & 92.8 & 44.4 & 75.4 & 83.8 \\ \hline
& \multicolumn{12}{c}{Coarse-Grained} \\
$\text{ADV}_{\text{64bit}}$ & \multicolumn{1}{c}{-} & \multicolumn{1}{c}{-} & \multicolumn{1}{c}{-} & \multicolumn{1}{c}{-} & \multicolumn{1}{c}{-} & \multicolumn{1}{c}{-} & 33.4 & 63.6 & 75.0 & 23.5 & 49.0 & 60.5 \\
UCCH & 6.2 & 17.9 & 26.5 & 4.4 & 14.9 & 24.5 & 12.7 & 30.7 & 40.6 & 8.7 & 25.8 & 37.4 \\
DADH & 0.4 & 2.3 & 4.7 & 0.3 & 1.5 & 3.0 & 0.1 & 0.7 & 2.6 & 0.2 & 0.8 & 1.6 \\ \hline
\end{tabular}
\caption{Comparison of instance-level retrieval performance across various models on MS-COCO and Flickr30K datasets. Results are reported as Recall@k (k=1,5,10) for both (i $\rightarrow$ t) and (t $\rightarrow$ i) tasks. Models are grouped into three categories: Vision-Language Pre-trained Fine-grained (VLP FG) models, non-pretrained Fine-Grained (FG) models, and Coarse-Grained (CG) models. BLIP-2 and X-VLM are additionally evaluated in a no-fusion (NF) configuration. ADV was only evaluated on Flickr30K due to the required model-specific dataset preprocessing files for MS-COCO being unavailable}
\label{tab:recall}
\end{table*}

\begin{table}[h]
\centering
\begin{tabular}{lllllll}  \hline
\multicolumn{1}{c}{\multirow{3}{*}{\begin{tabular}[c]{@{}c@{}}Instance\\  Retrieval\\  Recall@k:\end{tabular}}} & \multicolumn{6}{c}{Flickr30K} \\
\multicolumn{1}{c}{} & \multicolumn{3}{c}{Image to Text} & \multicolumn{3}{c}{Text to Image} \\
\multicolumn{1}{c}{} & \multicolumn{1}{c}{@50} & \multicolumn{1}{c}{@100} & \multicolumn{1}{c}{@200} & \multicolumn{1}{c}{@50} & \multicolumn{1}{c}{@100} & \multicolumn{1}{c}{@200} \\  \hline
 & \multicolumn{6}{c}{Coarse-Grained} \\
$\text{ADV}_{\text{64bit}}$ & 93 & 96.4 & 98.2 & 82.9 & 90.2 & 95.7 \\
UCCH & 68.4 & 81.3 & 88.5 & 68.7 & 80.6 & 89.5 \\
DADH & 8.6 & 16.3 & 28.0 & 8.3 & 15.4 & 26.3 \\ \hline
\end{tabular}
\caption{Extended top-k retrieval results (k=50, 100, 200) for \gls*{cg} models on the Flickr30K dataset}
\label{tab:recall2}
\end{table}

\textbf{Where \gls*{cg} succeeds.} The model ADV in its 64-bit \gls*{cg} setting achieves moderate success, with R@10 scores of 75.0\% for (i $\rightarrow$ t) and 60.5\% for (t $\rightarrow$ i) retrieval. The improvement of ADV over DADH and UCCH can primarily be attributed to adopting instance-level matching as the primary objective function. In its 2048-bit hash code \gls*{fg} setting, ADV achieves retrieval performance comparable to other continuous embedding-based \gls*{fg} models (Table \ref{tab:recall}). The extended top-k results in Table \ref{tab:recall2}---where $\text{ADV}_{\text{64bit}}$ attains R@100 scores of 96.4\% and 90.2\% for the (i $\rightarrow$ t) and (t $\rightarrow$ i) tasks---suggest that \gls*{cg} models have room to be used as initial retrieval candidate screening steps, provided the efficiency gained by employing such a strategy outweighs the information that is lost in this initial screening step.

\textbf{\gls*{cg} limitations.} The \gls*{cg} models DADH and UCCH were unable to properly capture instance-level relationships compared to the \gls*{fg} ones, as evidenced by their achieved R@1 scores (Table \ref{tab:recall}). Specifically, on the Flickr30K dataset, DADH achieves R@1 scores of 0.1\% for (i $\rightarrow$ t) and 0.2\% for (t $\rightarrow$ i) retrieval, while UCCH achieves R@1 scores of 12.7\% and 8.7\%, respectively. On the larger MS-COCO dataset, DADH achieves an R@10 score of 4.7\% for (i $\rightarrow$ t) and 3.0\% for (t $\rightarrow$ i) retrieval, whereas UCCH achieves R@10 scores of 26.5\%  and 24.5\%, respectively. This limited recall performance is primarily due to the objective function of these coarse models not targeting instance-level retrieval.

\textbf{Model comparison and attention mechanisms.} The recall performance difference between models employing query-time attention (BLIP-2, X-VLM, ViLT, IMRAM, SCAN) and those that do not is notable for its limited magnitude. In the case of pretrained models, the state-of-the-art fusion-reranking model BLIP-2 outperforms the dual-encoder model BEIT-3 by a moderate R@1 difference of 6.4 and 6.9 on the MS-COCO (i $\rightarrow$ t) and (t $\rightarrow$ i) tasks, respectively. The effectiveness of dual-encoder architectures is further evidenced by BEIT-3 achieving superior performance across all evaluation metrics compared to the full fusion-based ViLT, demonstrating that competitive retrieval performance can be attained without extensive inter-modal interaction at query time. Similarly, among non-pretrained models, the cross-attention model IMRAM outperforms the global-representation model VSRN by a modest R@1 difference of 4.2 for the MS-COCO (i $\rightarrow$ t) task and 1.6 for (t $\rightarrow$ i) task. Whether these moderate improvements in retrieval performance justify the use of query-time attention is a consideration that the experiments in Section \ref{sec:Experiment3} aim to inform.

\textbf{Dataset structure impact.} The dataset structure can explain the superior (i $\rightarrow$ t) retrieval performance across both datasets and all models compared to the (t $\rightarrow$ i) task. Each image query has five relevant captions to choose from, whereas each caption query has only one relevant image, making the (i $\rightarrow$ t) task inherently easier.

\subsection{Category-level retrieval results}
\begin{figure*}[htbp]
  \centering
  \begin{minipage}[b]{0.315\textwidth}
    \centering
    \includegraphics[width=\textwidth]{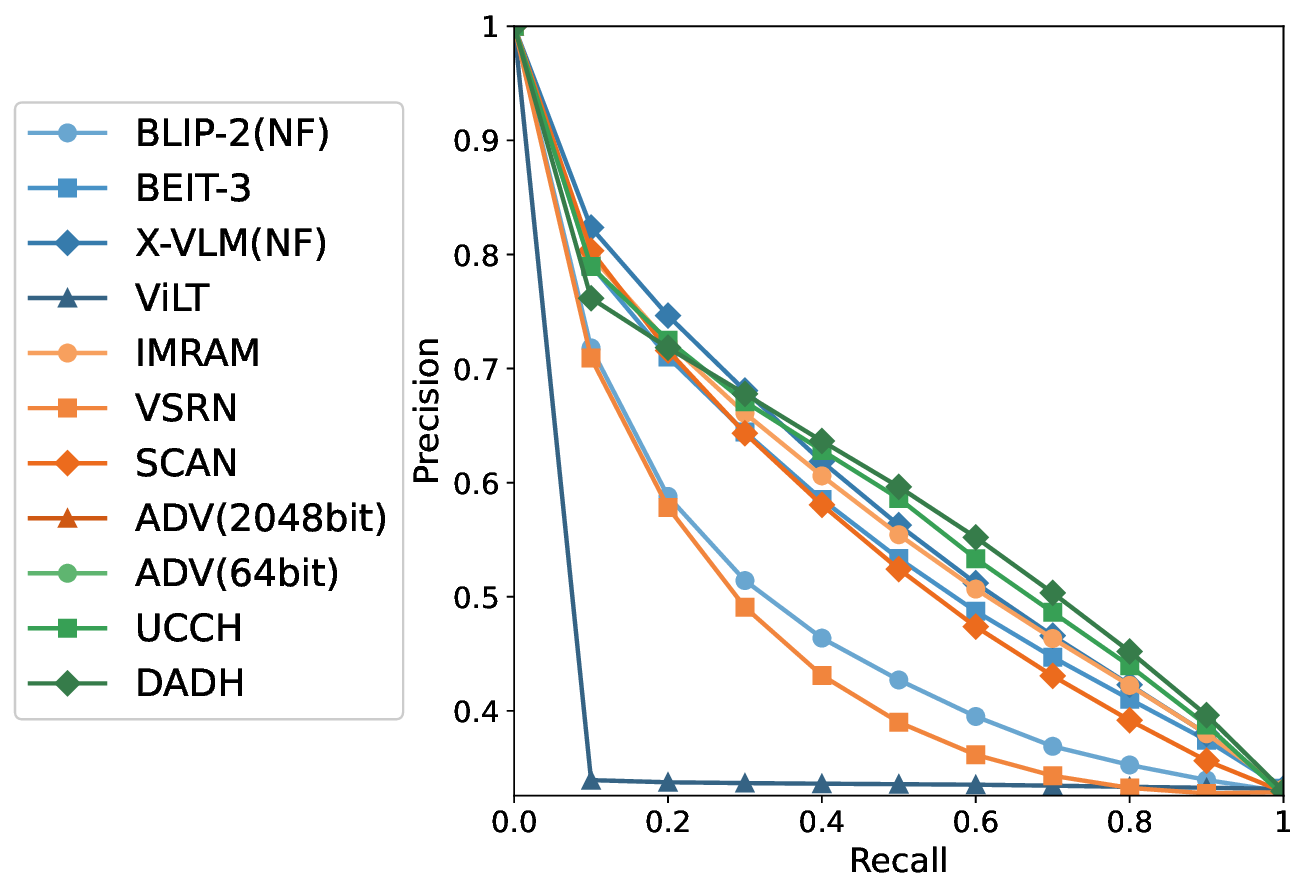}
    \subcaption{MS-COCO \\ Image $\rightarrow$ Text}
    \label{fig:subfigA}
  \end{minipage}
  \hfill
  \begin{minipage}[b]{0.22\textwidth}
    \centering
    \includegraphics[width=\textwidth]{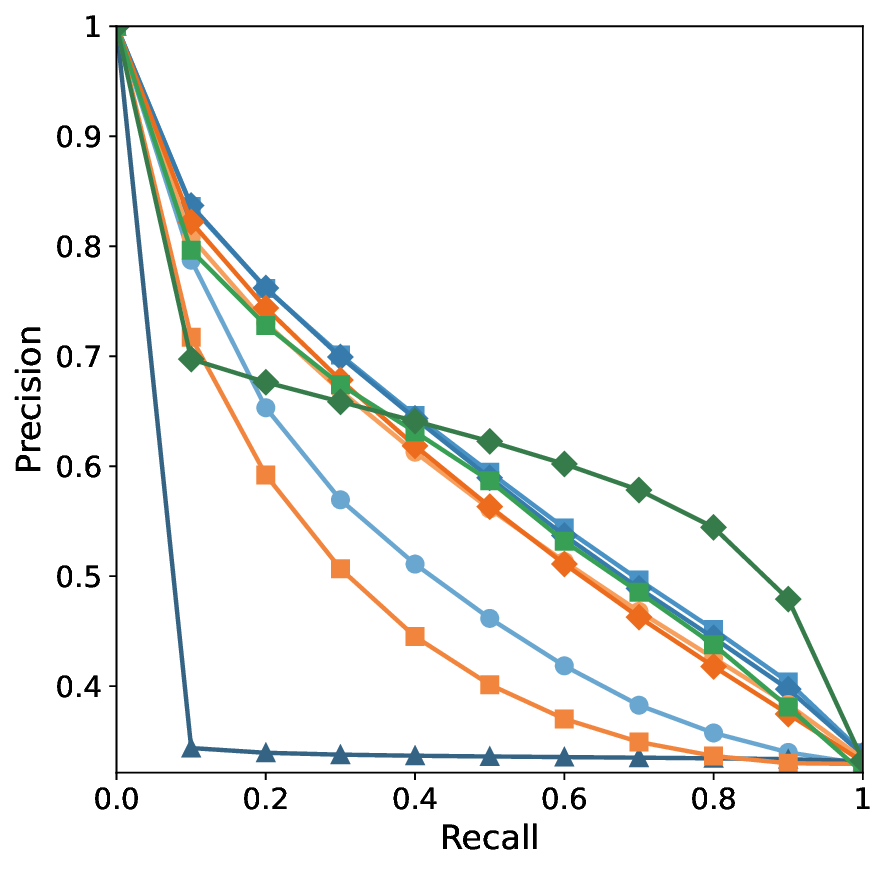}
    \subcaption{MS-COCO \\ Text $\rightarrow$ Image}
    \label{fig:subfigB}
  \end{minipage}
  \hfill
  \begin{minipage}[b]{0.22\textwidth}
    \centering
    \includegraphics[width=\textwidth]{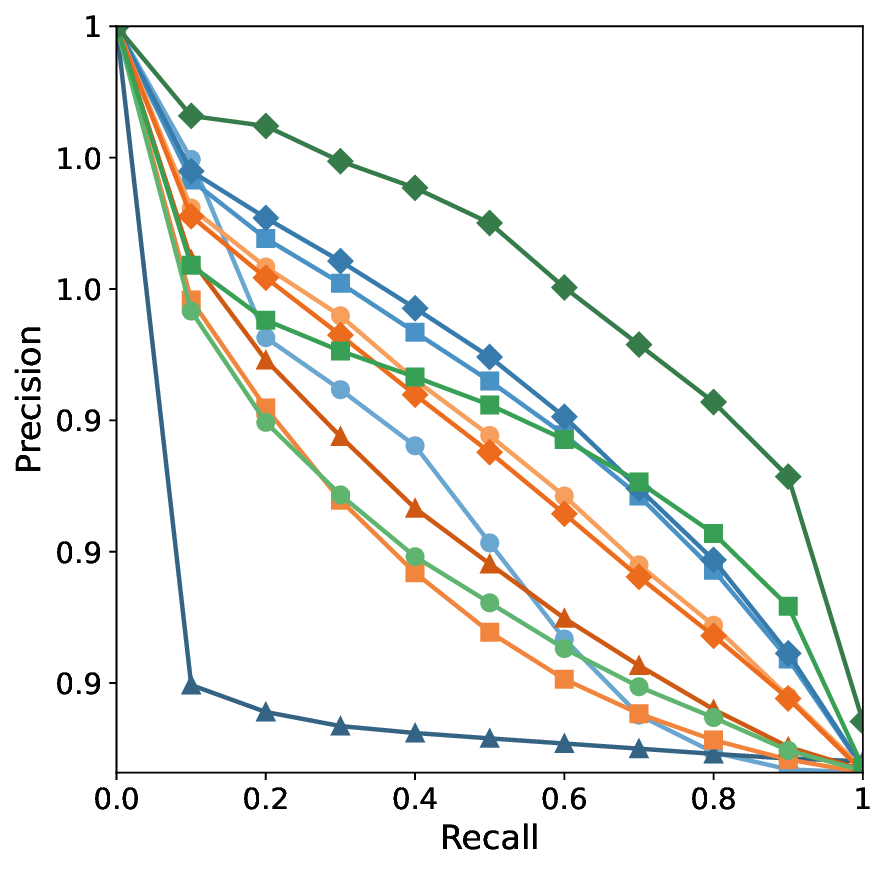}
    \subcaption{Flickr30K \\ Image $\rightarrow$ Text}
    \label{fig:subfigC}
  \end{minipage}
  \hfill
  \begin{minipage}[b]{0.22\textwidth}
    \centering
    \includegraphics[width=\textwidth]{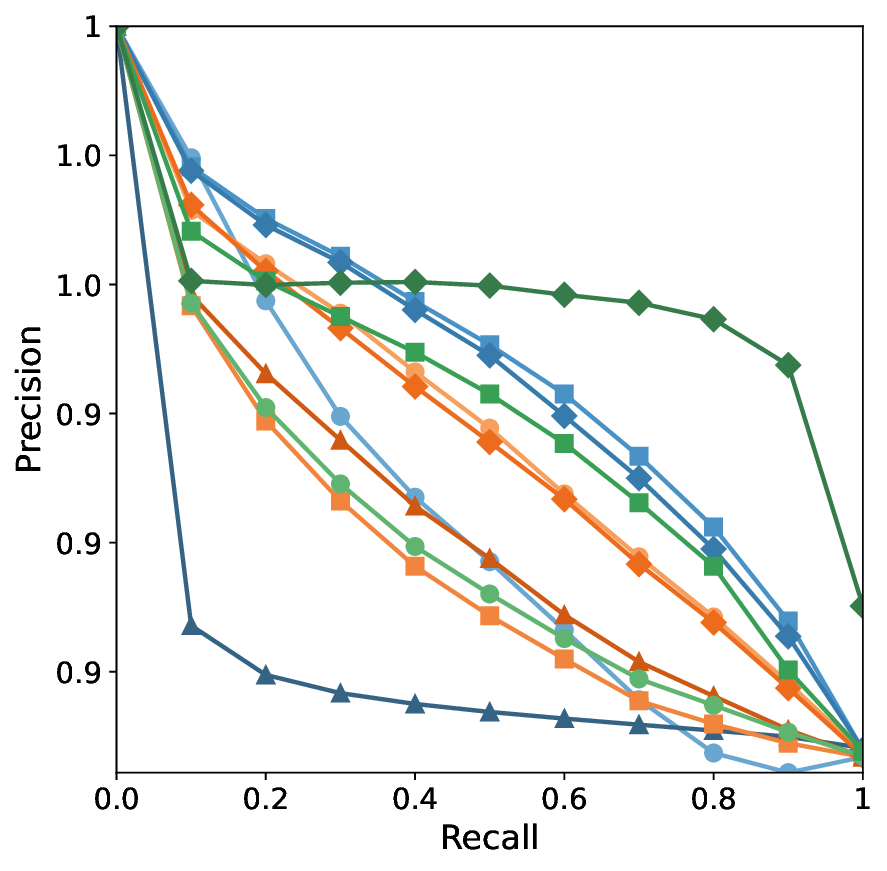}
    \subcaption{Flickr30K \\ Text $\rightarrow$ Image}
    \label{fig:subfigD}
  \end{minipage}

  \caption{Precision-recall curves of the selected models for the (i $\rightarrow$ t) and (t $\rightarrow$ i) tasks on the MS-COCO and Flickr30K datasets}
  \label{fig:categorypr}
\end{figure*}

This experiment aims to empirically compare \gls*{cg} and \gls*{fg} models in category-level retrieval tasks, where their relative performance is not well-established in the literature. The category-level mAP@k evaluation results for the selected models on the Flickr30K and MS-COCO datasets are reported in Table \ref{tab:map}. The 11-point interpolated precision-recall curves of the evaluated models for both datasets are shown in Figure \ref{fig:categorypr}. From these results, the following observations can be made:

\begin{table*}[ht]
\centering
\begin{tabular}{lllllllllllll} \hline
\multicolumn{13}{c}{Category-Level Retrieval Task}\\ \hline
\multicolumn{1}{c}{\multirow{3}{*}{\begin{tabular}[c]{@{}c@{}}\\ \\  mAP@k\end{tabular}}} & \multicolumn{6}{c}{MS-COCO} & \multicolumn{6}{c}{Flickr30K} \\
\multicolumn{1}{c}{} & \multicolumn{3}{c}{Image to Text} & \multicolumn{3}{c}{Text to Image} & \multicolumn{3}{c}{Image to Text} & \multicolumn{3}{c}{Text to Image mAP} \\
\multicolumn{1}{c}{} & \multicolumn{1}{c}{@10} & \multicolumn{1}{c}{@100} & \multicolumn{1}{c}{@N} & \multicolumn{1}{c}{@10} & \multicolumn{1}{c}{@100} & \multicolumn{1}{c}{@N} & \multicolumn{1}{c}{@10} & \multicolumn{1}{c}{@100} & \multicolumn{1}{c}{@N} & \multicolumn{1}{c}{@10} & \multicolumn{1}{c}{@100} & \multicolumn{1}{c}{@N} \\ \hline
& \multicolumn{12}{c}{Fine-Grained Vision-Language Pretrained} \\
BLIP-2 & \multicolumn{1}{r}{93.7} & \multicolumn{1}{r}{79.4} & n/a & \multicolumn{1}{r}{94.7} & \multicolumn{1}{r}{86.7} & \multicolumn{1}{r}{n/a} & 99.4 & 96.8 & n/a & 99.3 & 97.3 & n/a \\
$\text{BLIP-2}_{\text{NF}}$ & \textbf{96.2} & 87.8 & 47.9 & \textbf{95.9} & 88.5 & 51.1 & \textbf{99.5} & 98.1 & 92.5 & \textbf{99.5} & \textbf{98.2} & 92.5 \\
BEIT-3 & 95.4 & 84.8 & 56.1 & 95.6 & 88.3 & 60.7 & 99.3 & \textbf{97.5} & 94.3 & 99.2 & 97.5 & 94.6 \\
X-VLM & 95.0 & 86.1 & n/a & 94.6 & 83.4 & n/a & \textbf{99.5} & 96.7 & n/a & 99.0 & 96.9 & n/a \\
$\text{X-VLM}_{\text{NF}}$ & 95.8 & \textbf{87} & 58.4 & 95.8 & \textbf{88.8} & 60.3 & 99.3 & \textbf{97.7} & 94.5 & 99.3 & 97.6 & 94.5 \\
ViLT	& 63.5 &	42.6 &	32.9 &  62.8 &	42.5 &	33.0 &	97.2 &	92.0 &	88.7 &	94.6 &	90.1 &	88.7 \\ \hline
& \multicolumn{12}{c}{Fine-Grained} \\
IMRAM & 94.5 & 86.2 & 57.9 & 94.5 & 86.2 & 57.9 & 98.8 & 96.8 & 93.6 & 98.9 & 96.9 & 93.6 \\
VSRN & 95.1 & 83.5 & 45.8 & 94.7 & 83.5 & 46.6 & 98.8 & 95.7 & 91.6 & 98.8 & 95.7 & 91.6 \\
SCAN & 94.9 & 87.1 & 55.5 & 95.0 & 87.5 & 58.2 & 98.9 & 96.7 & 93.4 & 98.9 & 96.8 & 93.5 \\
$\text{ADV}_{\text{2048bit}}$ & \multicolumn{1}{c}{-} & \multicolumn{1}{c}{-} & \multicolumn{1}{c}{-} & \multicolumn{1}{c}{-} & \multicolumn{1}{c}{-} & \multicolumn{1}{c}{-} & 98.6 & 96.3 & 92.2 & 98.4 & 95.9 & 92.1 \\ \hline
 & \multicolumn{12}{c}{Coarse-Grained} \\
$\text{ADV}_{\text{64bit}}$ & \multicolumn{1}{c}{-} & \multicolumn{1}{c}{-} & \multicolumn{1}{c}{-} & \multicolumn{1}{c}{-} & \multicolumn{1}{c}{-} & \multicolumn{1}{c}{-} & 98.4 & 95.6 & 91.7 & 98.7 & 95.7 & 91.8 \\
UCCH & 86.2 & 81.5 & \textbf{60.7} & 86.6 & 82.1 & 60.2 & 95.1 & 92.1 & 89.3 & 90.7 & 88.2 & 88.8 \\
DADH & 84.8 & 78.4 & 58.7 & 75.6 & 71.6 & \textbf{60.5} & 98.6 & 97.8 & \textbf{96.3} & 96.7 & 96.1 & \textbf{95.7} \\ \hline
\end{tabular}
\caption{Comparison of category-level retrieval performance across various Vision-Language Pre-trained Fine-grained (VLP FG) models, non-pretrained Fine-Grained (FG) models, and Coarse-Grained (CG) models on the MS-COCO and Flickr30K datasets. Results are reported as mean Average Precision (mAP) at k (k=10,100,N) for both (i $\rightarrow$ t) and (t $\rightarrow$ i) tasks. Note that the fusion steps of BLIP-2 and X-VLM discard all samples that are left out of the fusion reranking step (k=128 and k=256 for Flickr30K and MS-COCO respectively). Therefore, mAP@N is not applicable (n/a) for these fusion-based models due to the number of relevant samples often exceeding the top-k of their reranking range. Hence, results for the no-fusion (NF) variants BLIP-2 and X-VLM are also provided}
\label{tab:map}
\end{table*}

\textbf{mAP@10 performance comparison.} Given the majority of the queries within both the Flickr30K and MS-COCO evaluation sets have hundreds to thousands of retrieval candidates which are considered relevant at the category level, scoring well on the mAP@10 metric is the least challenging aspect of this task. Nevertheless, the no fusion variant of BLIP-2 achieves an (i $\rightarrow$ t) mAP@10 score of 96.2\% and a (t $\rightarrow$ i) mAP@10 score of 95.9\% on the MS-COCO dataset, whereas the coarse model UCCH achieves scores of 86.2\% and 86.6\%, respectively (Table \ref{tab:map}). The superior performance of \gls*{fg} models on the mAP@10 metric can be attributed to instance-level matches being retrieved at the top ranks, which inherently share category labels with the query.

\textbf{mAP@N challenge and model size.} In contrast to the relative ease of scoring well on the mAP@10 metric, the mAP@N metric presents a more substantial challenge. Unlike instance-level retrieval, performance on this metric does not appear to align with the size of the model being used.  For example, on the MS-COCO dataset, the state-of-the-art \gls*{fg} \gls*{vlp} $\text{BLIP-2}_{\text{NF}}$ model achieves the second lowest category-level mAP@N score of 47.9\% on the (i $\rightarrow$ t) task, while the coarse model UCCH attains the highest score of 60.7\% (Table \ref{tab:map}). This suggests that \gls*{fg} low-level semantics may introduce noise when high-level category semantics is all that is needed for the task.

\textbf{Precision-recall trade-offs.} The precision-recall curves in Figure \ref{fig:categorypr} illustrate that the \gls*{cg} unsupervised model UCCH and, in particular, the supervised model DADH trained with learning objectives targeting category-level retrieval better maintain their precision as the proportion of relevant samples to be retrieved increases. In contrast, \gls*{fg} models, which are not optimised for retrieving large numbers of broadly relevant samples, show a sharper decline in precision with increases in recall requirement.

\textbf{Fusion-encoder global similarity limitation.} The fusion-encoder approach implemented by ViLT is less suitable for category-level retrieval due to both its architectural design and training approach. ViLT is trained using a binary matching approach where the model learns to distinguish correct pairings from a limited set of negative samples~\citep{kim2021vilt}. This setup trains the model to give high scores only to exact correct pairing while treating all other combinations as non-matches. However, for category-level retrieval, we need a model that can identify many relevant items that share semantic similarities with the query, not just find the single exact match. As a result, fusion-based models like ViLT, while effective at specific instance-level retrieval, perform poorly at ranking more generally similar items at the category level.

\textbf{Dataset difficulty comparison.} Across all models and tasks, the average mAP@N was 92.9\% for Flickr30K compared to 56\% for MS-COCO, indicating that the MS-COCO category-level retrieval benchmark presents a more challenging task. This difference in difficulty can be attributed to the datasets' composition. MS-COCO features more diverse and less overlapping labels, resulting in a more complex retrieval task at the category level. In contrast, Flickr30K has a narrower focus, with a majority of samples containing the `person' label due to its emphasis on human-object interactions. As a result, Flickr30K offers a more homogeneous set of images and labels, while MS-COCO provides a broader range of scenes and objects, leading to more varied queries and a more challenging test of model performance across diverse visual scenarios.

\subsection{Scaling encoding, storage, and attention} \label{sec:Experiment3}

This experiment aims to evaluate the scalability of the selected models by measuring encoding time and embedding storage cost across progressively larger test sets. We additionally measure query-time attention costs of applicable models to assess the trade-off in real-time performance which attention-based models face. We incrementally duplicate the Flickr30K evaluation set, generating four test sets: 1K/5K, 10K/50K, 100K/500K and 1M/5M image/text samples. By conducting these experiments on an NVIDIA A100 80GiB GPU and Intel Xeon Platinum 8480+ CPU, we aim to provide insights into the practical trade-offs between model complexity, computational resources, and retrieval performance at scale.

\textbf{Encoding time and storage cost.} Table \ref{tab:encoding} shows the encoding time and storage requirements for increasing data volumes for the selected models. The experiments were constrained by an 80GiB GPU memory threshold; runs exceeding this limit due to the size of the final or intermediate embeddings did not finish (DNF). For instance, BEIT-3, despite having the most compact final embeddings among VLP models, required 8.12GiB per 1K/5K images/captions to compute intermediate embeddings, causing its 10K/50K encoding run to exceed available GPU memory. These experiments used original model implementations without custom batching to ensure consistency and reflect practical limitations users might encounter with similar hardware. This consideration highlights the storage efficiency of cross-modal hashing models, which require only 0.36GiB to store 1M/5M image/text embeddings. For query-time attention models, Table \ref{tab:encoding} includes storage costs for original image and text features, which must be retained for the attention step; an important consideration when assessing the practicality of query-time attention in memory-constrained applications. Encoding time generally increased by a factor of 10 from \gls*{cg} to \gls*{fg} methods and again to VLP models, with X-VLM and ADV as exceptions. The rapid encoding times of UCCH and DADH demonstrate the computational efficiency of hash function encoders. ViLT's end-to-end fusion architecture computes similarity scores directly without encoding intermediate embeddings, requiring full model inference for each new query and preventing the possibility of offline embedding storage and reuse. Consequently, ViLT's computational costs are not represented in Table \ref{tab:encoding}, which focuses on one-time encoding costs for offline storage; instead, its query-time costs are analyzed in the following segment on attention mechanisms.

\begin{table*}[h]
\centering
\begin{tabular}{lllllll} \hline
 & \multicolumn{4}{c}{Encoding Time} & \multicolumn{2}{c}{Size (1M/5M)} \\
Img/Txt & 1K/5K & 10K/50K & 100K/500K & 1M/5M & Embed. & Features \\ \hline
& \multicolumn{6}{c}{Fine-Grained Vision-Language Pretrained} \\
BLIP-2 & 93.27s & 908.18s & DNF & DNF & 5.72GiB & 4240.25GiB \\
BEIT-3 & 56.76s & DNF & DNF & DNF & 17.2GiB & n/a \\
X-VLM & 12.42s & 100.87 s & DNF & DNF & 5.72GiB & 1126.82GiB \\
ViLT  &	n/a    & n/a	  & n/a	& n/a &	n/a     &	2723.37 GiB \\  \hline
& \multicolumn{6}{c}{Fine-Grained} \\
IMRAM & 50.4s & 47.35s & 458.10s & DNF & 45.75GiB & 3257.79GiB \\
VSRN & 5.03s & 19.55s & 180.08s & 1770.48s & 91.5GiB & n/a \\
SCAN & 6.47s & 51.7s & 464.464s & DNF & 45.75GiB & 3212.01GiB \\ \hline
 & \multicolumn{6}{c}{Coarse-Grained} \\
ADV 2048-Bit & 5.86s & 37.19s & 447s & 9854.41s & 11.71GiB & n/a \\
ADV 64-Bit & 5.75s & 35.29s & 334.67s & 3571.91s & 0.36GiB & n/a \\
UCCH & 0.40s & 4.12s & 60.57s & 488.28s & 0.36GiB & n/a \\
DADH & 0.53s & 1.13s & 6.46s & 59.41s & 0.36GiB & n/a \\ \hline
\end{tabular}
\caption{Encoding time cost for encoding 1K/5K, 10K/50K, 100K/500K and 1M/5M images/text in seconds and storage costs for holding 1M/5M image/text embeddings and features. ViLT's encoding time and embedding storage are not applicable due to its end-to-end fusion architecture, which computes similarity scores directly without encoding embeddings. For ViLT's computational costs, see Table \ref{tab:attntime}. DNF: Did not finish due to memory overflow. Features n/a: The model does not require original features during query time}
\label{tab:encoding}
\end{table*}

\textbf{Attention time.} Attention in this context refers to end-to-end fusion for ViLT, fusion-based reranking for BLIP-2 and X-VLM and cross-attention for IMRAM and SCAN; processes which are computed at query time. Due to the end-to-end fusion and cross-attention mechanisms used by ViLT, IMRAM and SCAN attending to all possible image-text pairs, the computational complexity of the mechanism is $O(n \times m)$, where n is the number of queries and m the number of retrieval candidates, quickly making retrieval prohibitively expensive computationally. This computational burden has driven the field towards dual-encoder architectures, which avoid query-time attention by computing and storing embeddings independently for each modality. For fusion-reranking, the time complexity is $O(n \times k)$, where k is the reranking shard threshold, which leads to improved costs relative to full attention. In the case of BLIP-2 and X-VLM, we kept $k = 128$ irrespective of the increase in sample size, which leads to a linear increase in attention computation time. However, if the k value were also scaled to the number of retrieval candidates, the computational complexity would approximate $O(n \times m)$. Such attention mechanisms, therefore, may be impractical for large-scale applications where query latency is critical.

\begin{table}[h]
\centering
\begin{tabular}{llll} \hline
        & \multicolumn{3}{c}{Attention (Images/Text)}          \\
        & 1K/5K   & 10K/50K            & 100K/500K \\ \hline
& \multicolumn{3}{c}{Transformer End-to-End Fusion} \\
ViLT    & 3093.76s   & >880h               & DNF     \\ \hline
& \multicolumn{3}{c}{Transformer Fusion Re-ranking} \\
BLIP-2  & 2376.89s & 24314.27s           & DNF     \\
X-VLM   & 370.48s  & 3707.37s            & DNF     \\ \hline
& \multicolumn{3}{c}{Cross-Attention} \\
IMRAM   & 190.67s  & 12200.28s           & DNF     \\
SCAN    & 17.51s   & 1285.34s            & DNF     \\ \hline
\end{tabular}
\caption{Time cost for computing end-to-end fusion, fusion re-ranking, cross-attention 1K/5K, 10K/50K and 100K/500K images/text. }
\label{tab:attntime}
\end{table}

\subsection{Scaling similarity search with FAISS }

The final step in retrieval, similarity search, is typically independent of the encoding model. Assuming no query-time attention or reranking, the model's task is complete once samples are encoded. These encoded samples are then typically passed to a separate specialised similarity search implementation. In that case, the only factors affecting similarity search time are the embedding type (real-valued continuous or bitwise hash code embeddings) and the embedding dimensions. To explore the practical implications of using \gls*{cg} binary hash codes compared to \gls*{fg} continuous embeddings, we employ Facebook AI Similarity Search (FAISS)~\citep{johnson21billion}, a robust and widely adopted~\citep{pizzi2022self, barrault2023seamlessm4t, thakur2021beir} similarity search implementation offering various indexes for both exhaustive search and Approximate Nearest Neighbour Search (ANNS). By using FAISS, we transition from model-specific comparisons to a generalised evaluation of embedding types within an industry-standard framework. The experiments were conducted using the following four indexes:

\textbf{Indexes.} IndexFlatIP is a continuous-embedding brute-force index that exhaustively searches the entire dataset via inner product and serves as our baseline for evaluating the performance of the other three indexes against it. IndexHNSW (Hierarchical Navigable Small World) is a graph-based index that organises vectors in a hierarchical structure and is the preferred index under FAISS guidelines for continuous embedding ANNS, given enough memory. IndexBinaryFlat is the exhaustive search index for binary vectors which computes the Hamming distance between all query and retrieval hash codes. IndexBinaryIVF (Inverted File System) is a partitioning-based index that divides the dataset into multiple inverted lists and is the preferred binary embedding ANNS index.

\textbf{Dataset.} The experiments were conducted using the $\text{SyntheticDataset}$ class provided by FAISS. The choice of embedding sizes within these experiments aimed to exploit the strengths of different granularity levels. For \gls*{cg} embeddings, a bit hash length of 64 was chosen to maximize efficiency based on previous experiments. For \gls*{fg} embeddings, a 2048-dimensional embedding, similar to VSRN, was selected to maximize accuracy.

\begin{figure}[htbp]
  \centering
  \begin{minipage}[b]{0.23\textwidth}
    \centering
    \includegraphics[width=\textwidth]{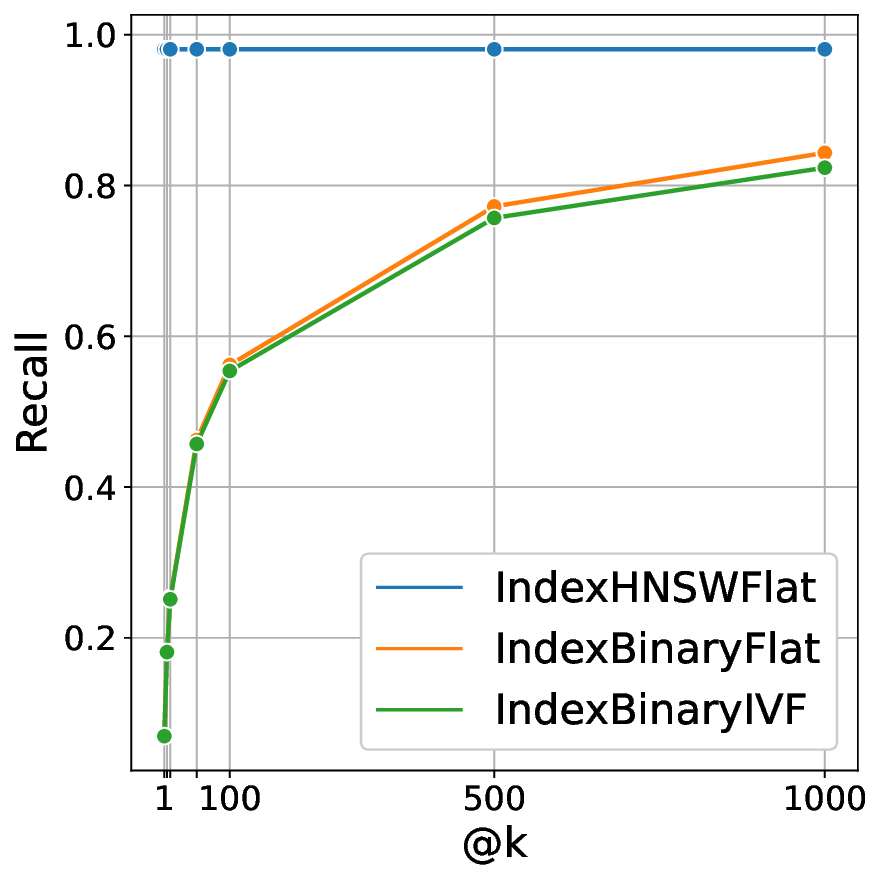}
    \subcaption{Retrieval performance}
    \label{fig:subfig2}
  \end{minipage}
  \hfill
  \begin{minipage}[b]{0.235\textwidth}
    \centering
    \includegraphics[width=\textwidth]{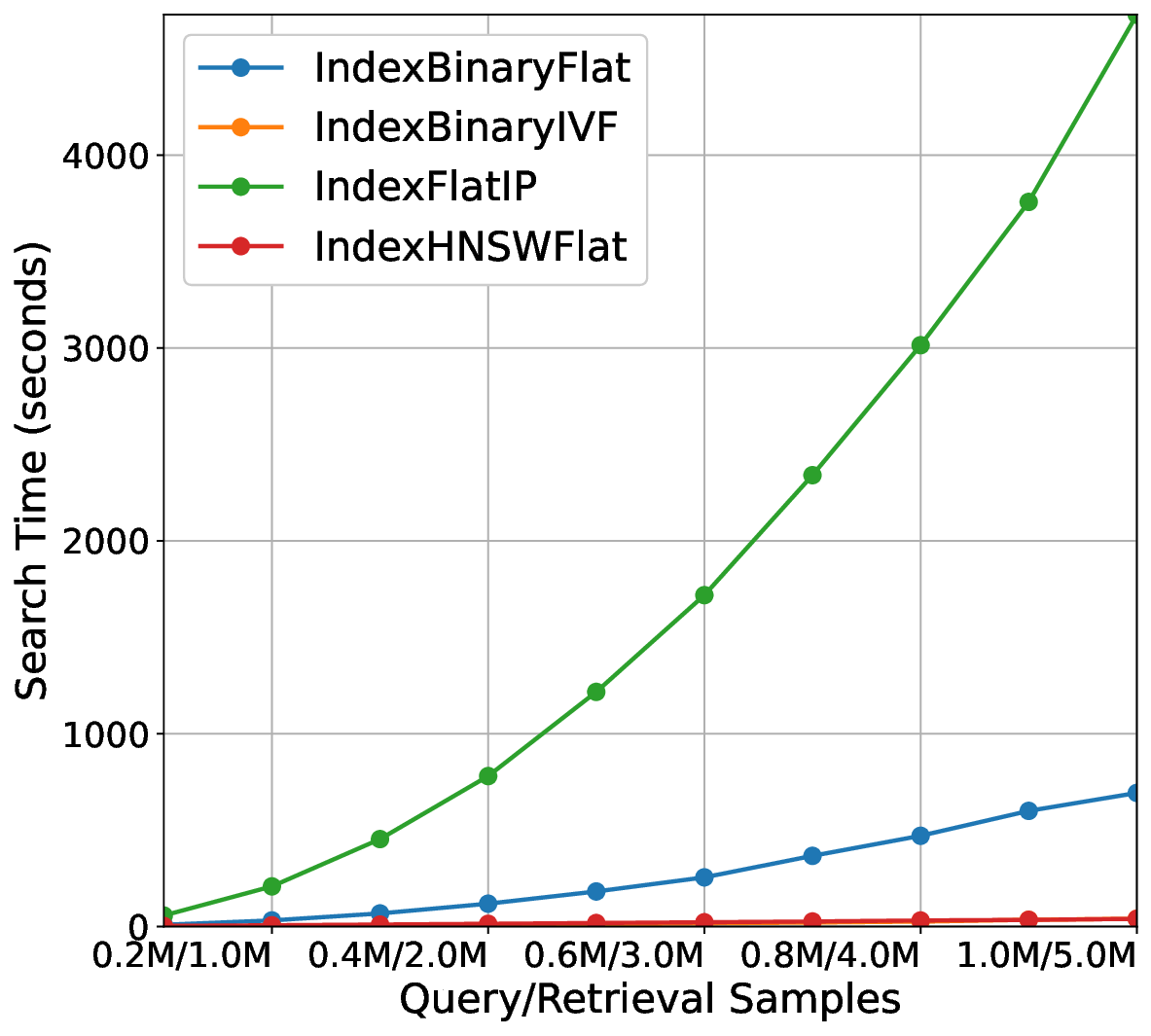}
    \subcaption{Search time}
    \label{fig:subfig1}
  \end{minipage}

  \caption{Comparison of search time and recall performance for both exhaustive and ANNS indexes of continuous and binary embeddings with increasing amounts of data using FAISS}
  \label{fig:faisscombo}
\end{figure}

From Table \ref{tab:faiss} and Figure \ref{fig:faisscombo}, the following observations can be made:

\begin{table*}[h]
\centering
\begin{tabular}{llllll} \hline
 & \multicolumn{5}{c}{100K Queries/500K Candidates} \\
 & R@1 & R@100 & R@1000 & Index & Search \\ \hline
$\text{FG Flat}_\text{2048D}$ & \multicolumn{3}{c}{Baseline (100.0)} & 1.93s & 57.17s \\
$\text{FG HNSW}_\text{2048D}$ & 99.2 & 99.2 & 99.2 & 32.44s & 3.09s \\
$\text{CG Flat}_\text{64-bit}$ & 10.7 & 60.6 & 93.6 & \textless{}0.01s & 8.84s \\
$\text{CG IVF}_\text{64-bit}$ & 10.5 & 68.9 & 89.7 & 0.33s & 0.70s \\  \hline
 & \multicolumn{5}{c}{500K Queries/2.5M Candidates} \\
 & R@1 & R@100 & R@1000 & Index & Search \\  \hline
$\text{FG Flat}_\text{2048D}$ & \multicolumn{3}{c}{Baseline (100.0)} & 9.55s & 1216.75s \\
$\text{FG HNSW}_\text{2048D}$ & 98.3 & 98.3 & 98.3 & 154.29s & 18.17s \\
$\text{CG Flat}_\text{64-bit}$ & 8.2 & 62.6 & 89.3 & \textless{}0.01s & 181.83s \\
$\text{CG IVF}_\text{64-bit}$ & 8.3 & 61.6 & 86.7 & 1.41s & 11.13s \\  \hline
 & \multicolumn{5}{c}{1M Queries/5M Candidates} \\
 & R@1 & R@100 & R@1000 & Index & Search \\  \hline
$\text{FG Flat}_\text{2048D}$ & \multicolumn{3}{c}{Baseline (100.0)} & 19.04s & 4728.18s \\
$\text{FG HNSW}_\text{2048D}$ & 98.0 & 98.0 & 98.0 & 300.65s & 39.80s \\
$\text{CG Flat}_\text{64-bit}$ & 6.9 & 56.1 & 84.3 & 0.02s & 692.80s \\
$\text{CG IVF}_\text{64-bit}$ & 6.9 & 55.3 & 82.3 & 2.82s & 40.20s \\  \hline
\end{tabular}
\caption{Recall and search time performance of Fine-Grained (FG) and Coarse-Grained (CG) FAISS indexes on the FAISS Synthetic Dataset. The FG exhaustive search (FG Flat) is the baseline ground truth to measure the scalability of other indexes relative to it. The time for `Index' refers to the construction time of the index. The time for `Search' refers to the search time for the entire test set using the previously built index}
\label{tab:faiss}
\end{table*}

\textbf{Index recall performance.} The HNSW index demonstrates high recall performance, maintaining 98\% of the recall relative to an exhaustive search in the 1M queries/5M retrieval candidates test case across all reported R@k scores (Table \ref{tab:faiss}). In contrast, binarising the embeddings to \gls*{cg} representations significantly impacts performance: For the 1M/5M test case, the \gls*{cg} approach retains only 6.9\% of R@1 performance compared to the exhaustive \gls*{fg} search. However, this retention improves at higher recall levels, reaching 84.3\% for R@1000 (Figure \ref{fig:faisscombo}). After the initial performance drop due to binarisation, further indexing these \gls*{cg} embeddings using an IVF index results in minimal additional recall degradation. For instance, in the 1M/5M test, the R@1 performance remains at 6.9\% for both flat and IVF indexes, while R@1000 shows a moderate decrease from 84.3\% to 82.3\%.

\textbf{Search scalability and hardware impact.} As visualised in Figure \ref{fig:faisscombo}, when not applying any ANNS indexes, the \gls*{cg} exhaustive search IndexBinaryFlat is a more scalable search approach than the \gls*{fg} IndexFlatIP. However, once the two embedding types are indexed by their preferred ANNS method - HNSW for \gls*{fg} embeddings and IVF for \gls*{cg} embeddings - their search times become comparable (Table \ref{tab:faiss}). For example, in the 1M/5M test, the search time for the \gls*{fg} HNSW (39.80s) is similar to the \gls*{cg} IVF (40.25s). Despite the potential for computational efficiency offered by \gls*{cg} methods through lightweight bitwise operations, Figure \ref{fig:faisscombo} illustrates a practical overlap in search time between \gls*{fg} and \gls*{cg} indexes across increasing data sizes. This convergence in search time can be attributed to recent hardware optimisations favouring continuous embeddings. This result suggests that binarised embeddings have room for further optimisation research efforts, particularly in standardising methodologies that optimise hardware utilisation to handle bitwise embeddings more efficiently.

\section{Key findings and discussion}\label{sec:discussion}

Based on the results obtained from the conducted comparative experiments, the following are the main key findings and recommendations derived from this study:

\textbf{Performance comparison.} \gls*{fg} models consistently outperformed \gls*{cg} models in instance-level retrieval tasks. \gls*{cg} models, however, demonstrated competitive performance in category-level retrieval, especially when retrieving large numbers of relevant samples. These findings challenge the conventional notion of \gls*{fg} models having universal superiority in retrieval performance over \gls*{cg} models; while \gls*{fg} models excel in specific instance-level retrieval, their performance advantage diminishes in broader, category-level tasks. This suggests that, when retrieval performance is the main concern, the choice between \gls*{fg} and \gls*{cg} models should be task-dependent rather than always defaulting to \gls*{fg} models.

\textbf{Hybrid coarse-to-fine potential.} An intuitive integration of \gls*{cg} and \gls*{fg} models could involve a two-step approach, where a \gls*{cg} model selects top-k candidates to reduce the computational load of a subsequent fine-grained reranking step. Our evaluation results showed that \gls*{cg} models trained specifically with instance-level loss functions could potentially serve as this initial screening step within a coarse-to-fine \gls*{itr} pipeline. However, traditional \gls*{cg} models trained on category-level loss functions are not suitable for this purpose. Despite their effectiveness in broadly retrieving samples of the same category as the query, these models do not consistently rank the exact instance-level match high enough in the retrieval rank to serve as an effective initial screening step.

\textbf{Attention mechanisms.} State-of-the-art benchmark recall results are achieved by models such as BLIP-2, which employ fusion-encoder reranking to refine search results. However, the computational load associated with such query-time attention mechanisms is difficult to justify for practical use in retrieval applications. This inefficiency stems from fusion encoders attending to every possible image-text pair with a computational complexity of $O(n \times m)$, where n is the number of queries and m is the number of retrieval candidates. Unlike retrieval tasks where fusion encoders must process all query-candidate pairs, in applications like VQA or image captioning, the fusion encoder only needs to attend to a single image-text pair at a time (e.g. the given question and image), resulting in $O(n)$ complexity. This makes fusion encoders particularly well-suited for these single-pair reasoning tasks\citep{ishmam2024image, stefanini2022show}, but prohibitively expensive for large-scale retrieval. Therefore, for retrieval applications where storage concerns can be addressed, dual-encoder VLP models which do not have query-time attention represent the most practical architecture type for instance-level retrieval.

\textbf{Search scalability insights.} Scalability experiments revealed that the potential efficiency advantages of \gls*{cg} models, particularly in terms of bitwise operations, do not always translate into practical performance gains for query latency. This was evidenced when applying FAISS-based ANNS indexing, where query latency of \gls*{cg} and \gls*{fg} embeddings was equalised, yet \gls*{fg} embeddings achieved considerably higher recall performance. This outcome highlights the significant impact of hardware optimisations and similarity search implementations on search performance. Without custom implementations of hardware-level optimised bitwise operations, \gls*{cg} embeddings do not improve query-time latency over \gls*{fg} ones when using a standard similarity search implementation such as FAISS. Further research into standardised optimisations of \gls*{cg} search for modern hardware architectures is recommended to fully leverage their potential efficiency advantages.

\textbf{Storage efficiency.} In terms of storage efficiency, \gls*{cg} models offer significant advantages. The 64-bit embeddings used for the evaluated \gls*{cg} models were over 15.8 times smaller than the most compact \gls*{fg} embeddings evaluated (256D). For category-level retrieval tasks with large amounts of relevant retrieval samples where \gls*{fg} models do not yield improvements in retrieval performance, the storage savings of \gls*{cg} models become particularly compelling.

\subsection{Limitations}

Our proposed \texttt{FiCo-ITR} library and experiments identify new insights into the tradeoffs of \gls*{fg} and CG ITR models, however, we acknowledge certain limitations. \gls*{cg} models are commonly benchmarked on datasets which contain hashtag collections as text samples (e.g. MIR-Flickr25K~\citep{huiskes2008mir} or NUS-WIDE~\citep{chua2009nus}). In contrast, \gls*{fg} models require full sentences to enable instance-level retrieval. There are currently no benchmark ITR datasets that contain both full sentences and hashtag collections as their text samples. This limitation presents an opportunity for an extension of our work, where an image captioning model could be used to generate full sentences for MIR-Flickr25K or NUS-WIDE to enable further evaluations.

Our scalability experiments quantified the computational costs associated with increasingly large retrieval sets. To achieve this, we used data duplication to grow our datasets. However, the potential degradation of retrieval performance as the size of the retrieval set increases could not be measured through this approach. Large-scale \gls*{itr} benchmarks which would enable this investigation do not exist in the field. Currently available large-scale image-text datasets (e.g. LAION-5B~\citep{schuhmann2022laion}, CC13~\citep{changpinyo2021conceptual}) are built for vision-language pretraining and have data imbalance and noise, which makes them unsuitable for benchmarking purposes. Our future work will target creating large-scale ITR benchmark datasets which maintain data quality and balance.

Our study could be extended in several directions: Our experiments focused on a cross-methodology comparison of \gls*{fg} and \gls*{cg} approaches. However, further within-methodology experiments could inform future work, such as implementing \gls*{fg} model ensembles or identifying the most suitable \gls*{cg} architecture for real-world use cases. Additional experiments could examine the generalisability of our findings across different data distributions, application domains, and model sensitivity to different types of data. Further efficiency gains could be achieved through hardware optimisations and custom batching solutions.

\section{Conclusion} \label{sec:conclusion}

While the conceptual differences in performance between \gls*{fg} and \gls*{cg} image-text retrieval models are well-documented in the literature, empirical data quantifying these differences has been sparse. This study introduced the library and toolkit \texttt{FiCo-ITR}, which provides a standardised toolkit for evaluating both \gls*{fg} and \gls*{cg} image-text retrieval models, which addresses this empirical gap. The results within this paper indicate that while \gls*{fg} methods excel in instance-level retrieval, \gls*{cg} approaches demonstrate competitive performance in category-level tasks where a large number of relevant samples must be retrieved. Despite the potential for computational efficiency offered by \gls*{cg} models, practical evaluations showed comparable query latencies between \gls*{fg} continuous embeddings and \gls*{cg} bitwise hash code embeddings, highlighting the practical impact of recent hardware optimisations. These results highlight that the notion of \gls*{fg} models offering more robust retrieval performance while \gls*{cg} models are more efficient is not straightforward; instead, these characteristics depend on the specific retrieval task and implementation. Future work will focus on two key areas: First, the development of a large-scale image-text retrieval benchmark to enable scalability experiments on real data. Second, the exploration of hybrid coarse-to-fine approaches that leverage insights from this study to balance retrieval performance and efficiency. These directions aim to further bridge the gap between \gls*{fg} and \gls*{cg} methodologies, potentially leading to more robust and scalable image-text retrieval systems.

\section*{Declarations}

\textbf{Code availability.} The source code for the \texttt{FiCo-ITR} library and toolkit can be found in the project's GitHub repository: \url{https://github.com/MikelWL/FiCo-ITR}. \\

\noindent \textbf{Conflict of interest.} The authors declare no Conflict of interest. \\

\noindent \textbf{Ethical approval.} This article contains no data or other information from studies or experimentation involving human or animal subjects.

\bibliography{sn-article}

\end{document}